\documentclass[aps,superscriptaddress,showkeys,showpacs,floatfix,amsmath,amssymb,twocolumn]{revtex4}
\usepackage{graphicx}
\usepackage{epsfig}
\usepackage{dcolumn}
\usepackage{bm}

\renewcommand{\r}{{\bf r}}

\newcommand{\be}{\begin{eqnarray}}
\newcommand{\ee}{\end{eqnarray}}

\begin{document}
\title{Global investigation of odd-even mass differences and radii with 
isospin dependent pairing interactions}

\author{C. A. Bertulani}
\author{Hongliang Liu}
\affiliation{Department of Physics, Texas A\&M University-Commerce,
Commerce, Texas 75429, USA}
\author{H. Sagawa}
\affiliation{Center for Mathematics and Physics, University of Aizu,
Aizu-Wakamatsu, 965-8580 Fukushima, Japan}\date{\today}

\begin{abstract}
The neutron and proton  odd-even mass differences are  systematically 
studied with
Hartree-Fock+BCS (HFBCS) calculations with Skyrme interactions  and
an isospin dependent contact pairing interaction.  The  strength of pairing 
interactions is determined to reproduce empirical odd-even mass differences 
in a wide region of mass table.  By using the optimal parameter, we perform 
global HF+BCS calculations of  nuclei and compare with experimental data.
The importance of isospin dependence of the pairing interaction is singled out for odd-even mass 
differences in 
medium and heavy isotopes. The proton and neutron radii are studied systematically by
using the same model.  
\end{abstract}


\maketitle

\section{Introduction}
Microscopic theories for calculating nuclear masses and/or binding energies (see, e.g., \cite{boh58,BM69,BB05}), have been revived and further elaborated with the advance of computational 
resources. These advances are now  sufficient to perform global studies based on, e.g., self-consistent mean field theory, sometimes also denoted by density functional theory (DFT) \cite{Ben03,Stoi06}.   
One particular aspect of the nuclear binding problem is a phenomenon of odd-even staggering (OES) 
of the binding energy.  Numerous microscopic calculations have been published that treat individual isotope 
chains.  However, it might be necessary to examine the whole body of OES data to draw
 general conclusions \cite{Satula98}. 
 
 Theoretically,  OES values are often inferred from the average HFBCS or Hartree-Fock-Bogoliubov gaps  \cite{dug01,mar07,Yama08},  
 rather than  directly calculated from the experimental binding energy differences  between 
 even and  odd nuclei. 
  It should be mentioned that the average HFB gaps are sometimes
 substantially different from the odd-even mass differences calculated from experimental binding energies.
 In this work,  we compare directly the calculated OES with the ones extracted from experiment.
One should say that there are also several prescriptions to obtain the OES from experiments, such as
 3-point, 4-point, and 5-point formulas  \cite{Satula98}.  We adopt the 3-point formula
  $\Delta^{(3)}$  centered at an odd nucleus, i.e., odd-N nucleus for neutron gap and
 odd-Z nucleus for proton gap \cite{BM69}:
\be  \label{eq:oes} 
\Delta^{(3)}(N,Z)\equiv\frac{\pi_{A+1}}{2} \Big[\mathrm{B}(N-1,Z)&-&2\mathrm{B}(N,Z)   \\\nonumber
&+&\mathrm{B}(N+1,Z)\Big] \; ,  \ee where $B(N,Z)$ is
the binding energy of $(N,Z)$ nucleus and
 $\pi_A=(-)^A$ is the number parity with $A=N+Z$.
 
For even nuclei, the OES is known to be sensitive not only to the
pairing gap, but also to mean field effects, i.e., shell effects
and deformations~\cite{Satula98,dug01}.  Therefore, the comparison of a
theoretical pairing gap with OES should be done with some discretion.
One advantage of $\Delta^{(3)}_o$ ($N=$ odd in Eq. \eqref{eq:oes})
is the suppression of the contributions from the mean field to the
gap energy.  Another advantage of $\Delta^{(3)}_o(N,Z)$ is that it can be applied to  more 
 experimental mass data than the higher order OES formulas. 
 At a shell closure, the OES~(Eq. \eqref{eq:oes}) does not go to zero
as expected, but it increases substantially. This large gap is an
artifact due to the shell effect, which is totally independent of
the pairing gap itself.

 Recently, an effective isospin dependent pairing interaction was proposed
  from the study of nuclear matter pairing gaps calculated by realistic
  nucleon-nucleon interactions.
  In Ref. \cite{mar07},
the density$-$dependent pairing interaction was defined as \be
V_{pair}(1,2)= \mathrm{V}_0 \,\mathrm{g}_\tau[\rho,\beta\tau_z]
\,\delta(\r_1-\r_2),
\label{eq:pairing_interaction} \ee
where $\rho=\rho_n+\rho_p$ is the
nuclear density and $\beta$ is the asymmetry parameter
 $\beta=(\rho_n -\rho_p)/\rho$.
The isovector dependence is introduced through  the
density-dependent term $\mathrm{g}_\tau$. The function
$\mathrm{g}_\tau$ is determined by the  pairing gaps in nuclear
matter and its functional form is given by
\be
\mathrm{g}_\tau[\rho,\beta\tau_z] =  1
-\mathrm{f}_\mathrm{s}(\beta\tau_z)\eta_\mathrm{s}
\left(\frac{\rho}{\rho_0}\right)^{\alpha_\mathrm{s}}
-\mathrm{f}_\mathrm{n}(\beta\tau_z)\eta_\mathrm{n}
\left(\frac{\rho}{\rho_0}\right)^{\alpha_\mathrm{n}},
\label{eq:g1t}
\ee where $\rho_0$=0.16~fm$^{-3}$ is the saturation
density of symmetric nuclear matter. We choose
$\mathrm{f}_\mathrm{s}(\beta\tau_z)=1-\mathrm{f}_\mathrm{n}(\beta\tau_z)$
and
$\mathrm{f}_\mathrm{n}(\beta\tau_z)=\beta\tau_z=\left[\rho_\mathrm{n}({\bf r})-\rho_\mathrm{p}({\bf r})
\right]\tau_z/\rho({\bf r})$. The parameters for $\mathrm{g}_\tau$ are obtained from the fit to the pairing gaps in
symmetric and neutron matter obtained by the microscopic
nucleon-nucleon interaction. 

\begin{figure}[htb]
\begin{center}
\includegraphics[clip,scale=0.33,angle=-90]{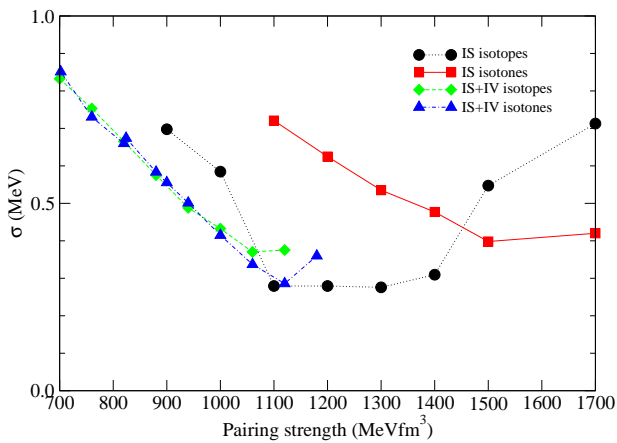}
\caption{(Color online) The mean square deviation $\sigma$ of OES between 
experimental data and the HF+BCS calculations.  The filled circles and squares correspond to 
the results with IS pairing for neutron and proton gaps, respectively, while 
 the filled diamonds and triangles are those of IS+IV pairing for neutron and proton gaps.
  Experimental data are taken from Ref. \cite{Audi2003}.  See the text for details. }
\label{fig01}
\end{center}
\end{figure}

\begin{table*}[htb]
\begin{center}
\setlength{\tabcolsep}{.06in}
\renewcommand{\arraystretch}{1.5}
\caption{Parameters for the density-dependent function
$\mathrm{g}_\tau$ defined in Eqs.~(\ref{eq:pairing_interaction}) and (\ref{eq:g1t}) for the 
  IS+IV interaction  (first row)
and $\mathrm{g}_s$ in Eq. \eqref{isoscalar} for the IS interaction.
The parameters for $\mathrm{g}_\tau$ are obtained from the fit to the pairing gaps in
symmetric and neutron matter obtained with the microscopic
nucleon-nucleon interaction. The paring strength $V_0$
  is adjusted to give the best fit
to odd-even staggering of nuclear masses. The parameters for $\mathrm{g}_s$
correspond to a surface peaked pairing interaction with no isospin
dependence. \label{table1}}
\begin{tabular}{cccccccc}
\toprule
interaction & $V_0$ (MeVfm$^3$)& $\rho_0$ (fm$^{-3}$) & $\eta_\mathrm{s}$ & $\alpha_\mathrm{s}$ & $\eta_\mathrm{n}$ & $\alpha_\mathrm{n}$ \\
\colrule
$\mathrm{g}_\tau$ (isotopes) & 1040  & 0.16 & 0.677& 0.365 & 0.931  & 0.378  \\
$\mathrm{g}_\tau$ (isotones)& 1120  & 0.16 & 0.677& 0.365 & 0.931  & 0.378  \\
$\mathrm{g}_s$ (isotopes)& 1300  & 0.16  & 1. & 1. & ---  & ---  \\
$\mathrm{g}_s$ (isotones)& 1500  & 0.16  & 1. & 1. & ---  & ---  \\
\botrule  
\end{tabular}
\end{center}

\end{table*}%

In the literature and in many mean field codes publicly available such as the original EV8 code \cite{EV8}, a pure contact interaction
is used without an isospin dependence. In our notation, this amounts replacing
the isospin dependent function $\mathrm{g}_\tau$ in Eq. \eqref{eq:pairing_interaction}  by the isoscalar
function
 \begin{equation}
\mathrm{g}_s=1-\eta_\mathrm{s}
\left(\frac{\rho}{\rho_0}\right)^{\alpha_\mathrm{s}}
\label{isoscalar}.
\end{equation}
The EV8 code has been modified, using the filling approximation,  to account for mass calculations for odd-N and odd-Z nuclei and is publicly available as the EV8odd code \cite{BB08}. It has also been modified to include isospin dependent paring, by means of Eq. \eqref{eq:pairing_interaction}.
 The parameters of the isoscalar interaction were adjusted with EV8 to a best global fit of nuclear masses \cite{bertsch09}. They correspond to
a surface peaked pairing interaction (Eq. \eqref{isoscalar} with $\eta_s$ not too far from the unity).

A recent publication has explored the isospin dependence of the pairing force for the OES effect for a few selected isotopic and isotonic chains \cite{BLS09}. Here we have made a more ambitious study by extending the calculation to the whole nuclear chart. We have also explored several other
 observables such as neutron and proton radii systematically which may allow  for   
more solid conclusions on isospin dependent pairing interactions.    

This paper is organized as follows. 
In section II we discuss our numerical calculation strategy. Our results are presented in section  III for the energies, separation energies, OES energies, and nuclear radii. Our conclusions are presented in section IV.

\section{Calculation strategy}

The HF+BCS calculations are performed by using SLy4 Skyrme interaction which was found to be the most accurate interaction for studying OES for a few selected ($N=50,82$) isotonic  and (Sn and Pb)isotopic chains  \cite{BLS09}.
Our iteration procedure used in connection to EV8odd
achieves an accuracy of about 100 keV, or less, with 500 Hartree-Fock iterations for each nuclear state. Our calculations were performed with the now decommissioned  XT4 Jaguar supercomputer at ORNL, as part of the UNEDF-SciDAC-2 collaboration \cite{SciDAC}.

The HF+BCS calculations were first performed for even-even nuclei.
The variables in the theory are the orbital wave functions $\phi_i$
and the BCS amplitudes $v_i$ and $u_i = \sqrt{1 - v^2_i}$. By
solving the BCS equations for the amplitudes, one obtains the
pairing energy from
\begin{equation}
E_{pair} = \sum_{i\neq j} V_{ij} u_i v_i u_j
v_j + \sum_i V_{ii} v^2_i \label{Epair}
\end{equation}
where $V_{ij}$ are the matrix elements
of the pairing interaction, Eq. \eqref{eq:pairing_interaction}, namely
$$
V_{ij}=V_0\int d^3 r |\phi_i({\bf r})|^2 |\phi_j({\bf r})|^2
\mathrm{g}_\tau[\rho ({\bf r}),\beta({\bf r})\tau_z],
$$
where $\rho({\bf r})=\sum_i v_i^2 |\phi_i({\bf r})|^2$.

\begin{figure*}[t]
\begin{center}
\includegraphics[clip,scale=0.42]{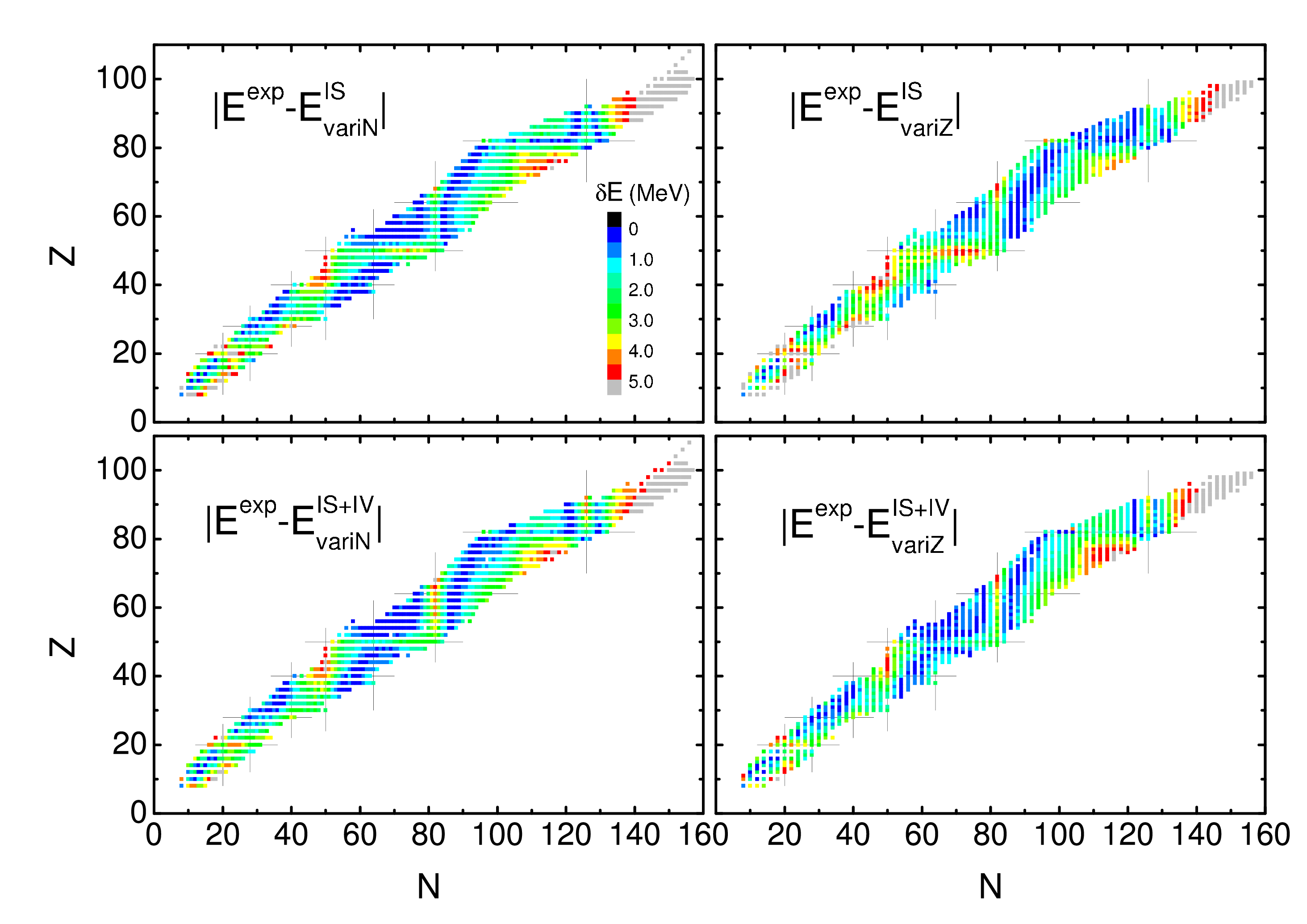}
\caption{(Color online) Binding energy differences between experimental data and calculations 
using the HF+BCS model with the IS and IS+IV pairing interactions.   The left  panels show the differences for even Z isotopes 
varying neutron numbers including both odd and even numbers.  
The right panel show those for even N isotones
varying proton numbers including both odd and even numbers. The thin lines show the closed 
shells at $N(Z)$ = 20, 28, 40, 50, 64, 82 and 126.
  Experimental data are taken from Ref. \cite{Audi2003}.  See the text for details. }
\label{fig-dE}
\end{center}
\end{figure*}

\begin{figure*}[htb]
\begin{center}
\includegraphics[clip,scale=0.42]{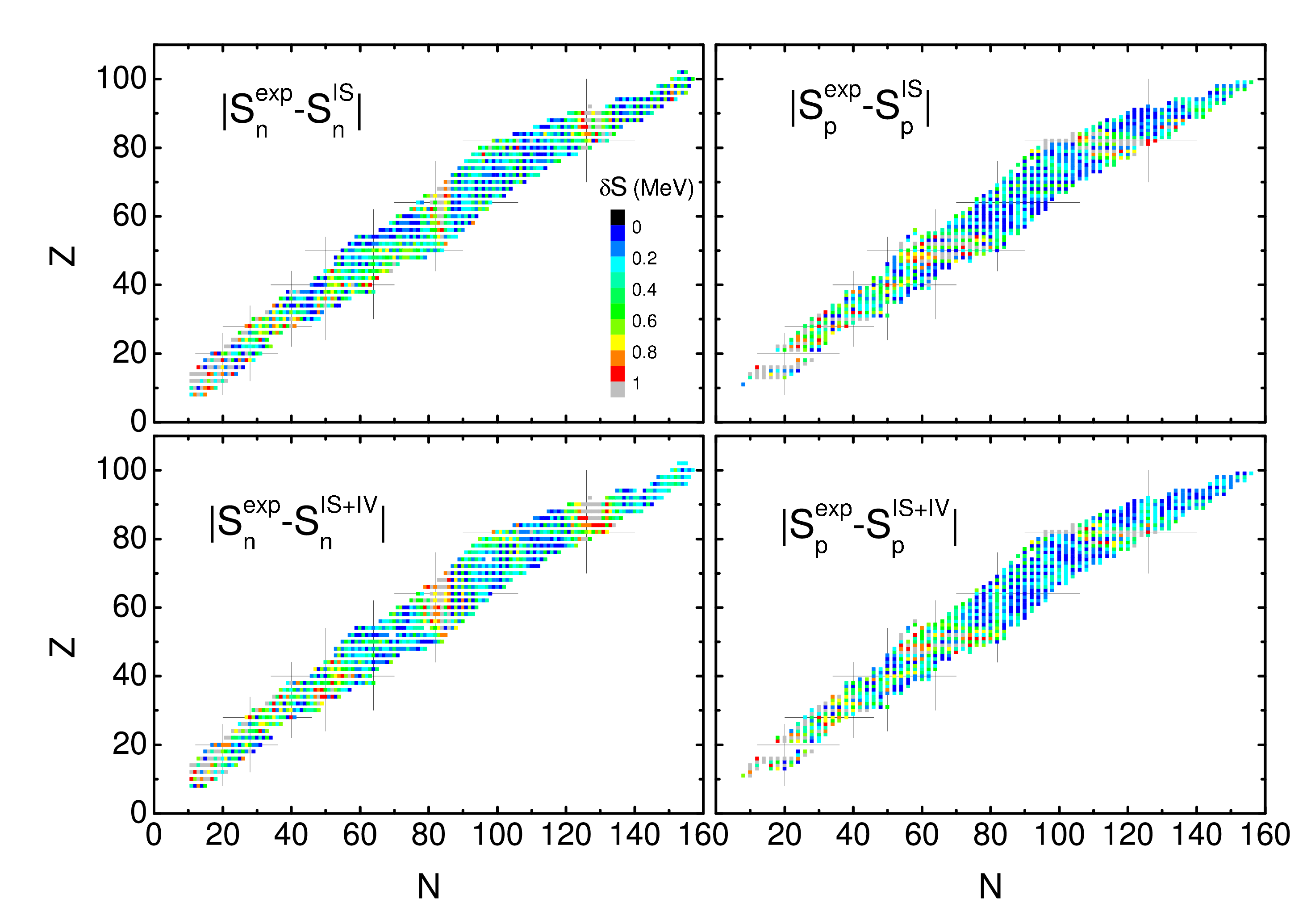}
\caption{(Color online)  The same as Fig. \ref{fig-dE} but for neutron and proton separation energies.  
See the caption to Fig. \ref{fig-dE} and the  text for details. }
\label{fig-dS}
\end{center}
\end{figure*}

After determining the single-particle energies of
even-even nuclei,
 the odd-A nuclei are calculated with the so-called filling
approximation for the odd particle starting from the HF+BCS
solutions of neighboring even-even nuclei: ones selects a pair of 
$i$ and $\widetilde{i}$ orbitals to be blocked, and changes the BCS parameters
 $v^2_i$ and  $v_{\widetilde{i}}^2$ for these  orbitals.
  The change is to set 
  $v^2_i = v_{\widetilde{i}}^2=1/2$  in
Eq. \eqref{Epair} for the pairing energy at an orbital near the Fermi
energy. Note that this  approximation gives equal occupation
numbers to both time-reversed partners, and does not account for the
effects of time-odd fields. More details of the procedure are
presented in Ref. \cite{bertsch09}.

The effect of 
  time-odd HF fields on the mass were studied in Refs. \cite{Duguet-Jerome,Mar11}.
It was pointed out that the effect of the time-odd fields is of the order of 100 keV
 for the binding energy 
 depending strongly on the configuration of the last particle,  and does not show 
 any clear sign of isospin dependence.  Thus the time-odd
  field  might not change   
 conclusions of the present study in the following,  while quantitative 
 accuracy might need some fine tuning of the pairing parameters.

\begin{figure*}[htb]
\begin{center}
\vspace{-0.5cm}
\includegraphics[clip,scale=0.6,angle=-90]{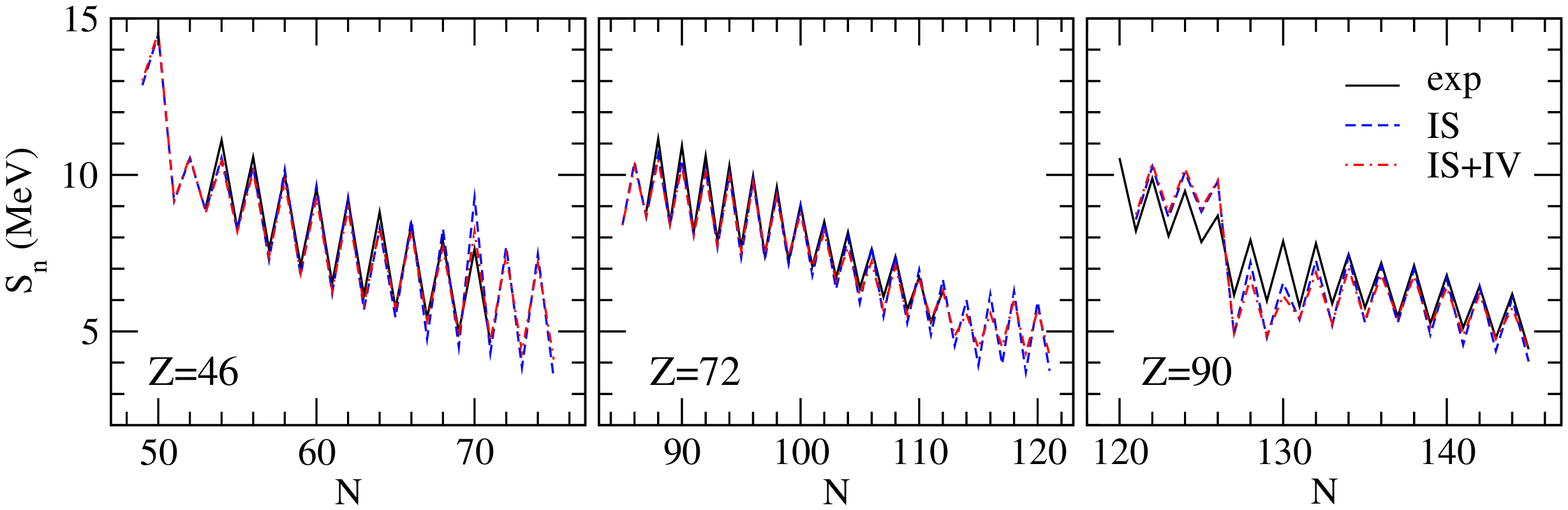}
\includegraphics[clip,scale=0.6,angle=-90]{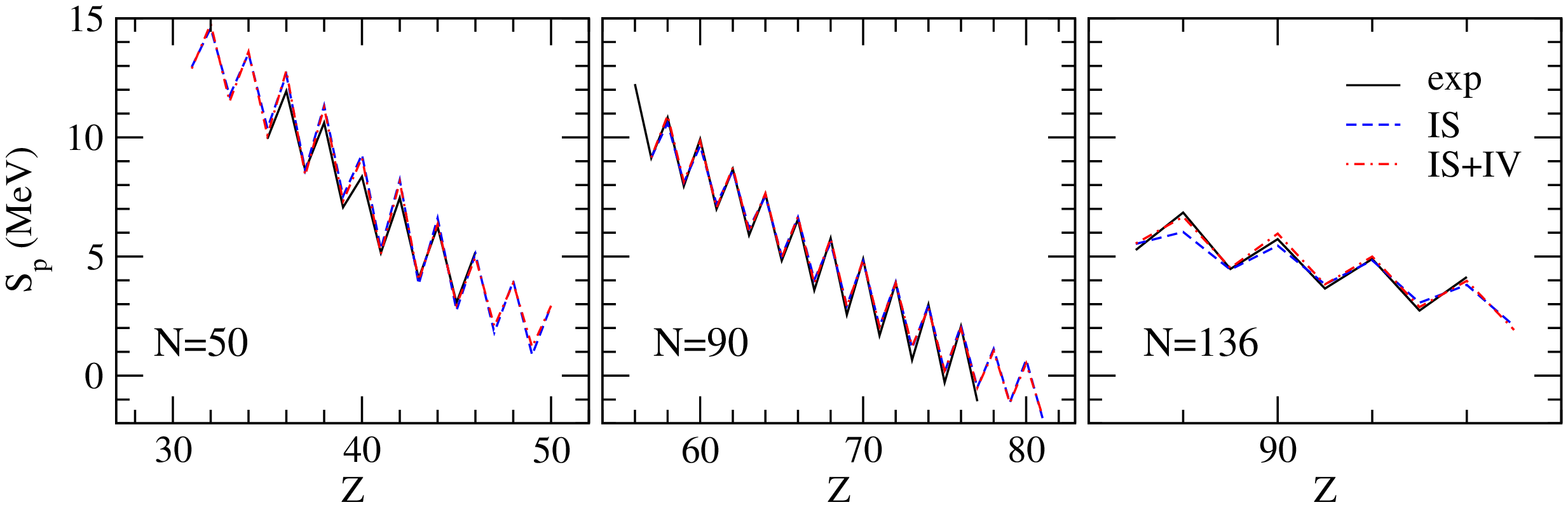}
\caption{(Color online) Neutron separation energies $S_n$
 of three isotope chains with $Z=46$, 72 and 90  calculated 
by IS and IS+IV interactions in HF+BCS model (upper panels). 
  The lower panels show the proton separation energies $S_p$ for $N=50$, 90 and 136 isotones. 
  Experimental data are taken from Ref. \cite{Audi2003}.  See the text for details. }
\label{fig-S}
\end{center}
\end{figure*}

\begin{figure}[htb]
\begin{center}
\includegraphics[clip,scale=0.45,angle=-90]{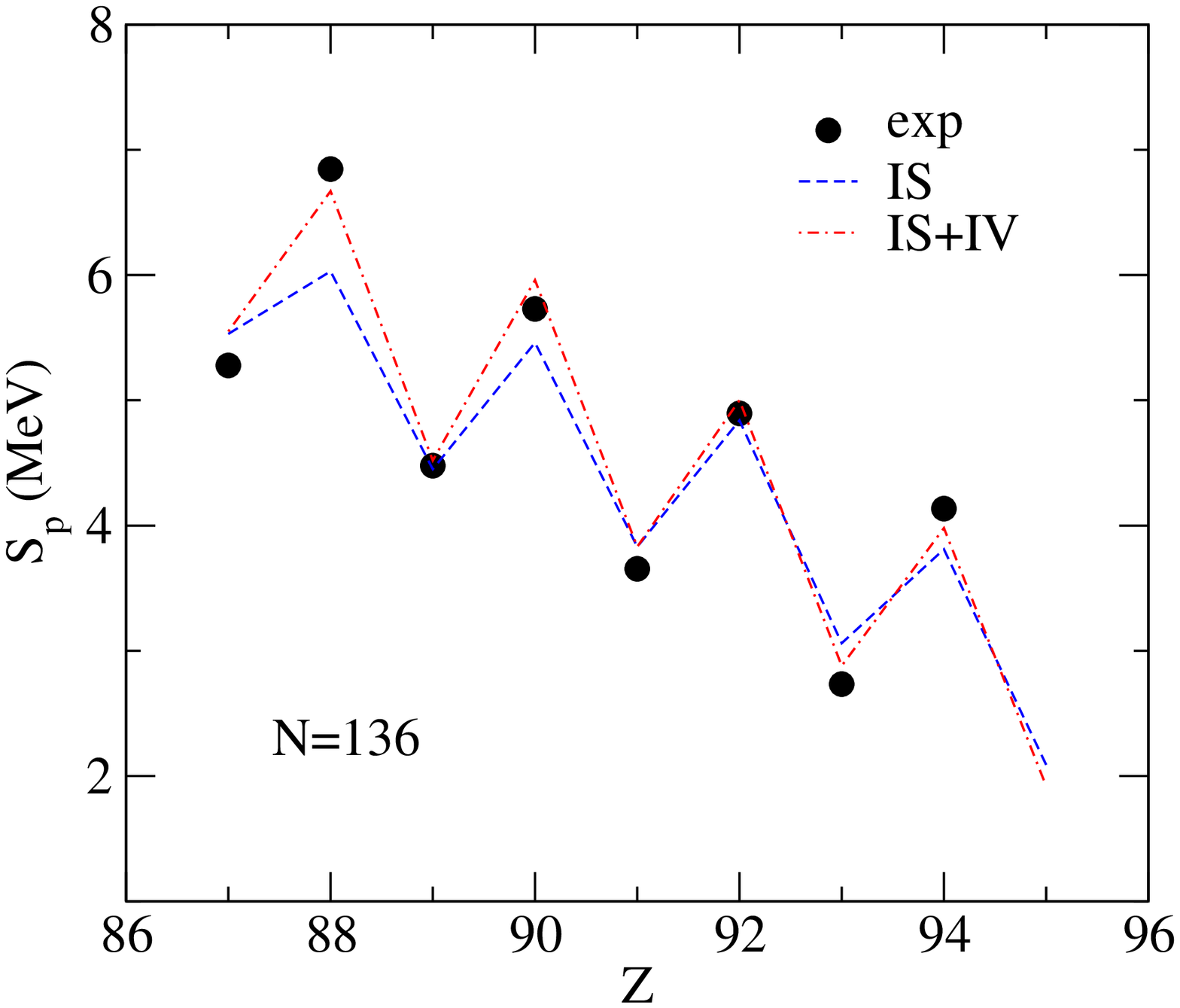}
\caption{(Color online) Proton  separation energies $S_p$
 of  $N=136$  isotones calculated 
by IS and IS+IV interactions in HF+BCS model.  
  Experimental data are taken from Ref. \cite{Audi2003}.  See the text for details. }
\label{fig-Sp}
\end{center}
\end{figure}

For the pairing channels we have taken the surface-type contact
interaction, Eq. \eqref{isoscalar}, and the isospin dependent
interaction, Eq. \eqref{eq:g1t}.  The density dependence of the
latter one is essentially the mixed-type interaction between the
surface and the volume types. The pairing strength $V_0$ depends on
the energy window adopted for BCS calculations. The odd nucleus is
treated in the filling approximation, by blocking one of the
orbitals. The blocking candidates are chosen within an energy window
of 10 MeV around the Fermi energy.
This energy window is rather small, but it is the maximum allowed by 
the program EV8odd.  It is shown that the BCS model used in the  EV8odd code gives 
almost  equivalent results 
 to the HF+Bogoliubov model with a larger energy window,
 except for unstable nuclei very close to the neutron drip line \cite{bertsch09}. 
The pairing strengths $V_0$ for IS and IS+IV pairing interactions are  
  adjusted to give the best fit
to odd-even staggering of nuclear masses in a wide region of the mass table.

\begin{figure*}[htb]
\begin{center}
\includegraphics[clip,scale=0.42]{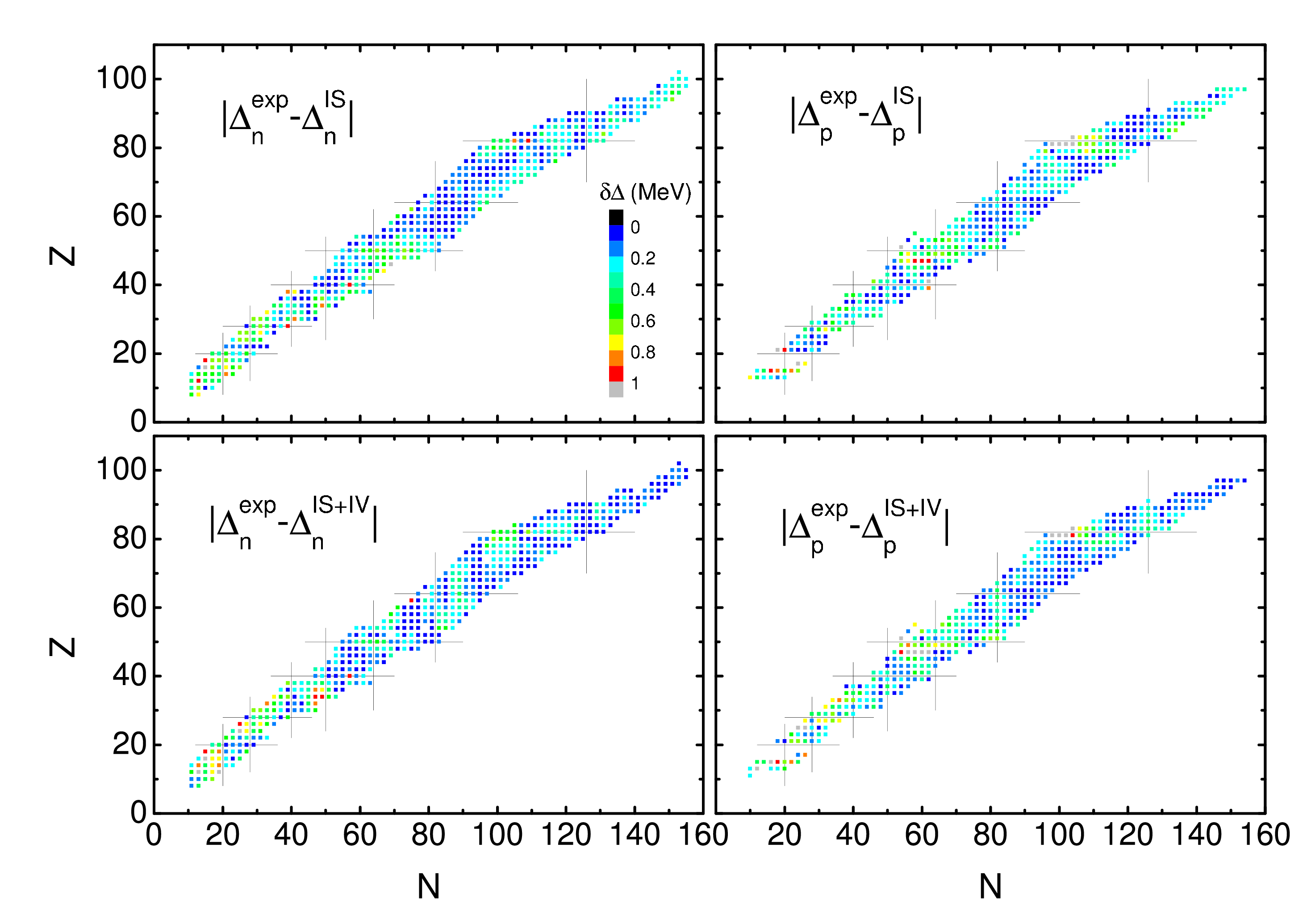}
\caption{(Color online) The same as in Fig. \ref{fig-dE} but for OES for neutrons and protons.
  See the caption to Fig. \ref{fig-dE} and the text for details. }
\label{figdD}
\end{center}
\end{figure*}

\section{Numerical results}
\subsection{Global data on odd-even staggering}

The results for the mean square deviation of our global mass table calculations are shown in Fig. \ref{fig01}.  
Table \ref{table1} gives the values $V_0$ in Eq. \eqref{eq:pairing_interaction} and the parameters for 
$\mathrm{g}_\tau$ and $\mathrm{g}_s$ is Eqs. \eqref{eq:g1t} and \eqref{isoscalar} used in the present work. 
Optimal pairing strength values were found to be different for isotones (varying $Z$, constant $N$) and
for isotopes  (varying $N$, constant $Z$).

Figure  \ref{fig01} shows  the mean  square deviation $\sigma$ of OES  between 
experimental data and the HF+BCS calculations.  The  mean  square deviation $\sigma$ is 
defined as
\be
\sigma =\sqrt{\sum_{i=1}^{N_i}\left|\Delta_i^{(3)}(HF+BCS)-\Delta_i^{(3)}(exp)\right|^2/N_i}
\ee
where $N_i$ is the number of data points.  
For the IS interaction, the results for neutrons show a shallow minimum at $V_0\sim(1100-1300)$ 
 MeV$\cdot$fm$^3$.  For protons, the minimum becomes at around  $V_0\sim1500$ MeV$\cdot$fm$^3$.
This difference makes it difficult to determine a unique pairing strength common for both neutrons and
protons.  The results of IS+IV pairing show a minimum at V$_0\sim1100$ MeV$\cdot$fm$^3$ for both 
neutron and proton OES which makes it easier to determine the value for the pairing strength. 
  Adopted values for the following calculations are listed in Table \ref{table1}.

The systematic study of HF+BCS calculations are performed for various isotopes and isotones 
for all available data sets with ($Z=8,\cdots,102$) and ($N=8,\cdots, 156$), respectively.  
Binding energy differences between experimental data and HF+BCS 
\be
\delta E=|E^{exp}-E^{cal}|
\label{Diff-E}
\ee
are shown for both IS and 
IS+IV pairing interactions in Fig. \ref{fig-dE}. The left panels show the values $\delta E$ 
varying neutron numbers (including both odd and even numbers) for each  even Z. 
The right panels show the values $\delta E$ 
 varying proton  numbers (including both odd and even numbers) for each  even N. 
 With IS pairing, we can see a rather large 
deviation for $Z=50$ isotopes  in the upper right panel. This difference disappears in the case 
of IS+IV pairing shown in the lower right panel.  On the other hand, for  the $N=82$ nuclei,
the IS+IV interaction does not work that well.  As far as the binding energies are concerned,
 the best results with IS+IV interaction are obtained for nuclei with $N=60-78$ and $N=86-96$. 

Separation energy differences between experimental data and HF+BCS for
 protons and neutrons are plotted in Fig.  \ref{fig-dS}.
The HF+BCS results of neutron separation energies $S_n$  are reasonable   
for medium and heavy mass nuclei with $N=60-120$, except 
near the closed shell $N=82$. For heavy nuclei with $N=126$, the calculated results are 
poorer than in other mass regions.   For proton separation energy $S_p$,  the HF+BCS also gives 
reasonable results, except in the $Z=50$ and 82 mass regions.

In order to see the different outcomes between IS and IS+IV pairing interactions, the HF+BCS model 
calculations are shown together with empirical data in Fig. \ref{fig-S}. 
In most of cases, the difference between the two pairing interactions are small. 
 However, we can see a clear improvement of the agreement  of $S_p$ 
with empirical data of $N=136$ isotones 
with IS+IV pairing  in Fig. \ref{fig-Sp}.  

\begin{figure*}[htb]
\begin{center}
\vspace{-0.5cm}
\includegraphics[clip,scale=0.6,angle=-90]{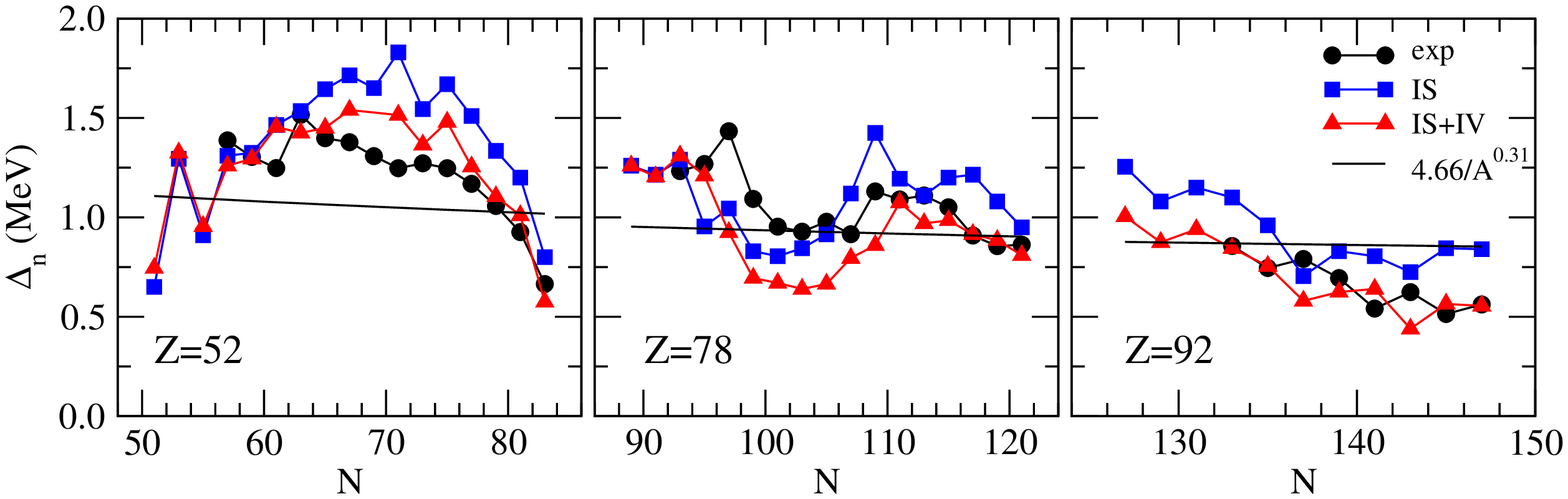}
\includegraphics[clip,scale=0.6,angle=-90]{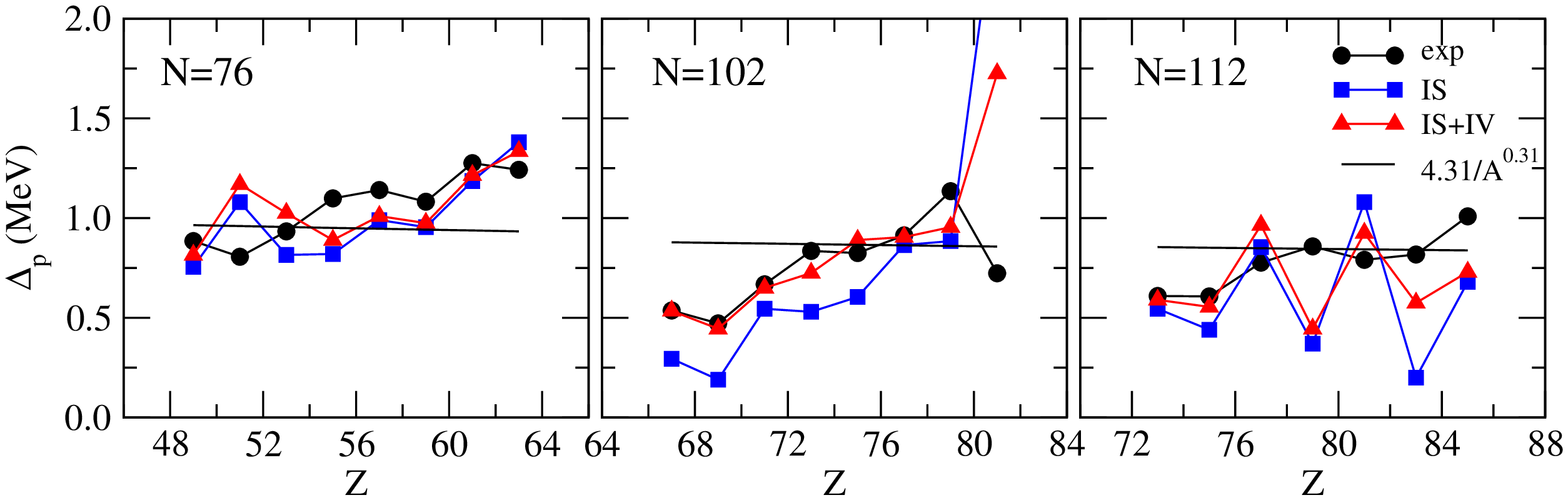}
\caption{(Color online) The  neutron and proton OES, $\Delta_n$ and  $\Delta_p$, calculated 
with the HF+BCS model with IS and IS+IV interactions.  
  Experimental data are taken from Ref. \cite{Audi2003}.  See the text for details. }
\label{fig-Delta-np}
\end{center}
\end{figure*}

The differences of neutron OES $\Delta_n$ and proton OES $\Delta_p$  
 between HF+BCS and empirical data are shown in 
Fig.  \ref{figdD} for both IS and IS+IV pairing, respectively, in the upper and lower 
panels. 
 The agreement between HF+BCS and the empirical data 
 are good in the overall mass region except for masses with $Z=50$ and 
at a small mass region $A<60$.   To clarify the difference between IS and  IS+IV pairing,
the OES differences $\Delta_n$ are shown in the upper panel of Fig. \ref{fig-Delta-np} 
for $Z=52$, 78 and 92 isotopes. The HF+BCS results
are compared with the experimental data and also the
phenomenological parameterization based on liquid drop model, 
\begin{equation}
 \bar{\Delta}=c/A^{\alpha}
\label{eq:gap-pheno}
\end{equation}
with $c=4.66(4.31)$ MeV for neutrons (protons)
and $\alpha=0.31$ which gives the rms residual of
0.25 MeV \cite{bertsch09}. 
 We can see clearly a better agreement of
IS+IV results with empirical data for all isotopes.  In the lower panel of Fig. \ref{fig-Delta-np},
 the OES differences $\Delta_n$ are shown for $N=76$, 102 and 112 isotones.
 The results with IS+IV pairing 
certainly improve systematically the agreement with empirical data, especially for $N=102$ isotones. 
The large increase of the HF+BCS model results at $Z=81$ is an artifact due to the shell 
closure at $Z=82$.  It is interesting to notice that the liquid drop formula gives smooth 
mass number dependence which reflects well that of very heavy isotones with $N=112$.  

\begin{table*}
\caption{\label{table2}Average $\Delta^{(3)}$ for low isospin and high isospin nuclei
 and its difference.  See the text for details.  }
\begin{ruledtabular}
\begin{tabular}{cccccc}
 \multicolumn{3}{c}{Data set} & Low isospin & High isospin & Difference \\
\hline
 Neutrons  & $Z=52$ & Exp   & 1.36 & 1.08 & -0.28 \\
           &        & IS    & 1.52 & 1.41 & -0.11 \\
           &        & IS+IV & 1.40 & 1.19 & -0.21 \\
           & $Z=78$ & Exp   & 1.13 & 0.99 & -0.14 \\
           &        & IS    & 0.96 & 1.16 &  0.20 \\
           &        & IS+IV & 0.87 & 0.91 &  0.04 \\
           & $Z=92$ & Exp   & 0.77 & 0.56 & -0.21 \\
           &        & IS    & 0.90 & 0.80 & -0.10 \\
           &        & IS+IV & 0.70 & 0.55 & -0.15 \\
 Protons   & $N=76$ & Exp   & 1.19 & 0.93 & -0.26 \\
           &        & IS    & 1.13 & 0.87 & -0.26 \\
           &        & IS+IV & 1.13 & 0.98 & -0.15 \\
           & $N=102$& Exp   & 0.96 & 0.63 & -0.33 \\
           &        & IS    & 0.79 & 0.39 & -0.40 \\
           &        & IS+IV & 0.92 & 0.59 & -0.33 \\
           & $Z=112$& Exp   & 0.87 & 0.66 & -0.21 \\
           &        & IS    & 0.58 & 0.61 &  0.03 \\
           &        & IS+IV & 0.67 & 0.70 &  0.03 \\
\end{tabular}
\end{ruledtabular}
\end{table*}

\begin{figure*}[htb]
\begin{center}
\includegraphics[clip,scale=0.4,angle=-90]{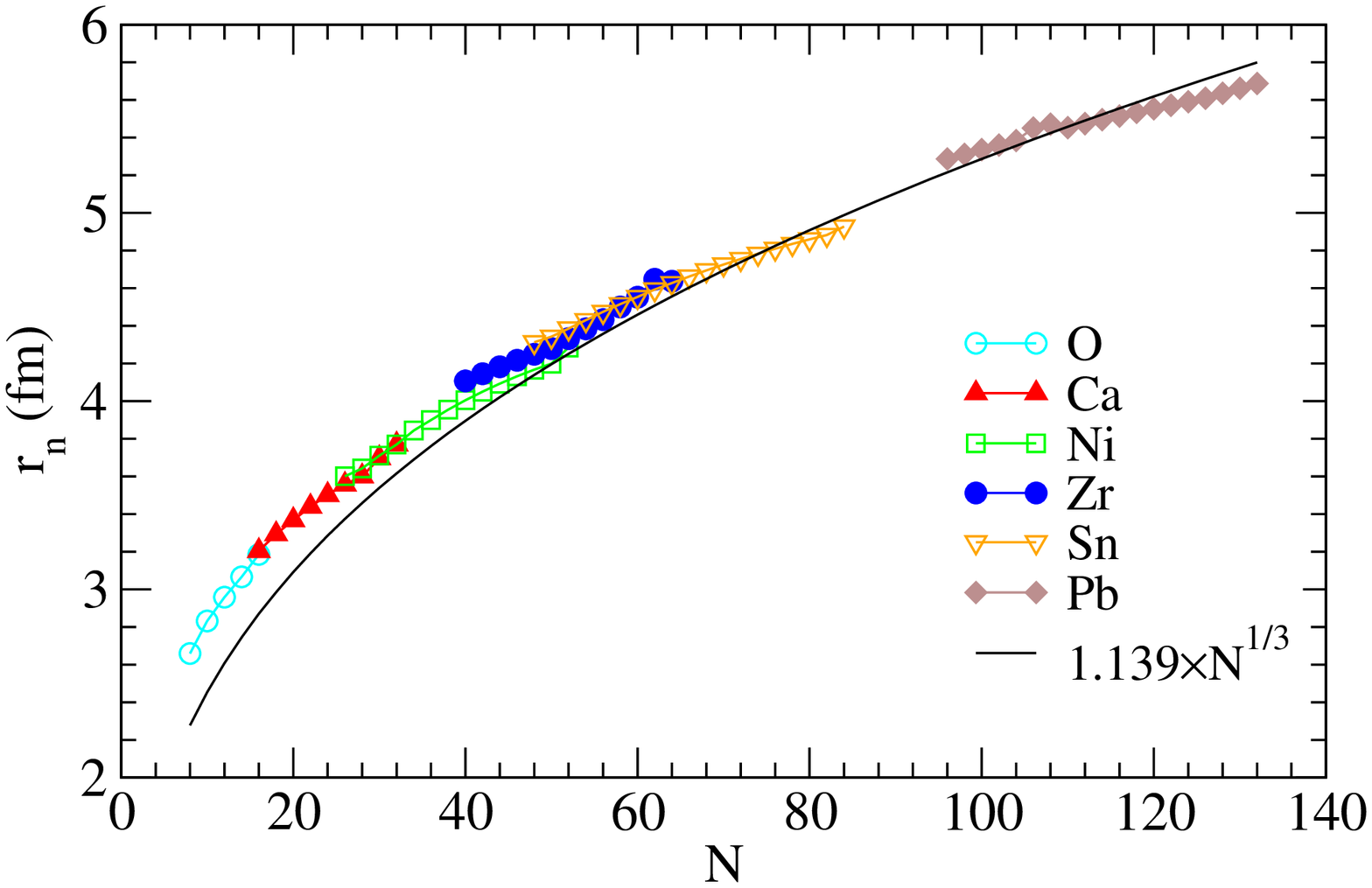}
\includegraphics[clip,scale=0.4,angle=-90]{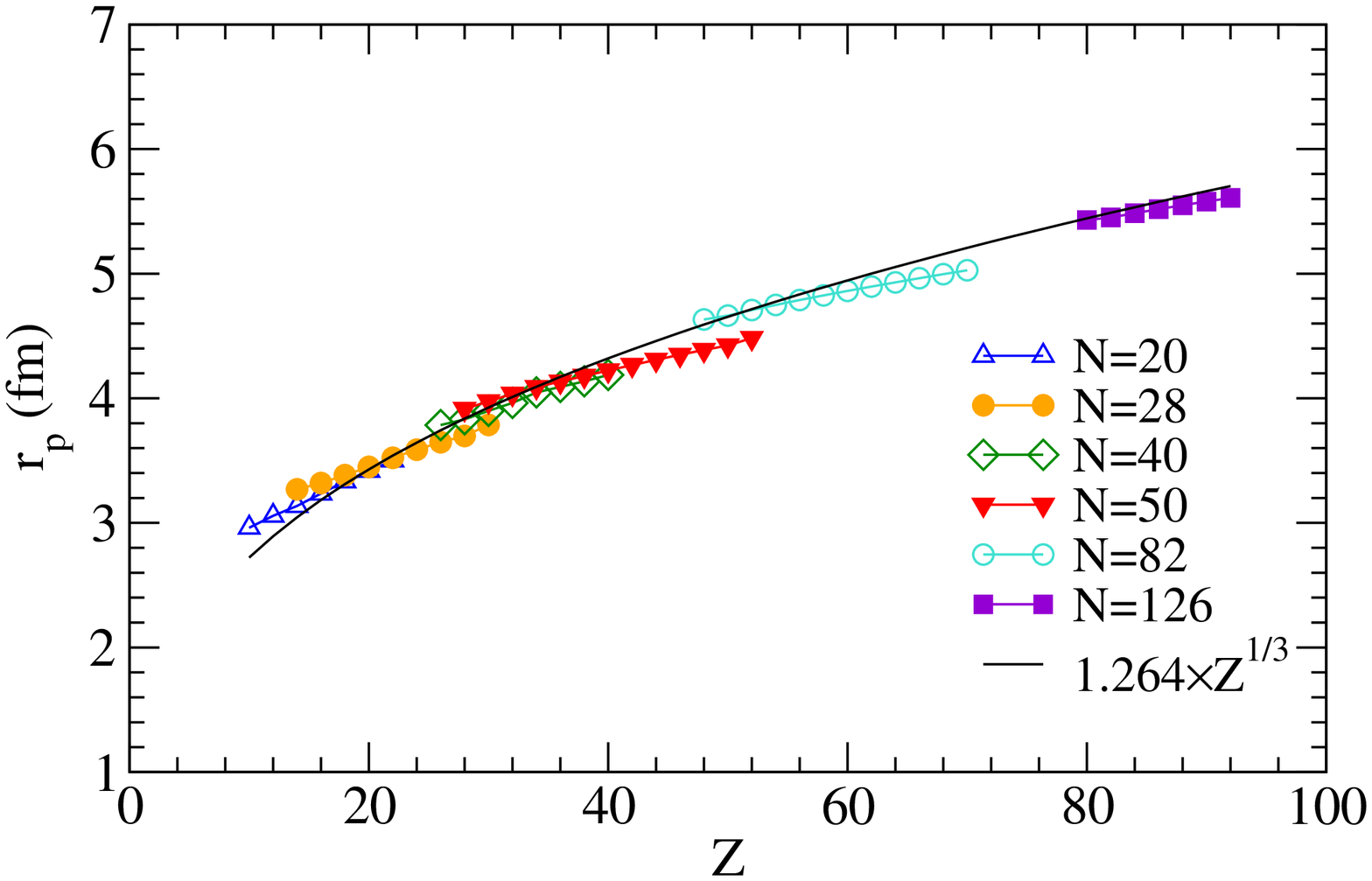}
\caption{(Color online) Neutron and proton radii of various isotopes and 
isotones calculated
by means of the HF+BCS model with  IS+IV interaction. The solid lines are empirical fits
 used in  Ref. \cite{Meng06}.  
See the text for details. }
\label{RnRp}
\end{center}
\end{figure*}

\begin{figure*}[htb]
\begin{center}
\includegraphics[clip,scale=0.6]{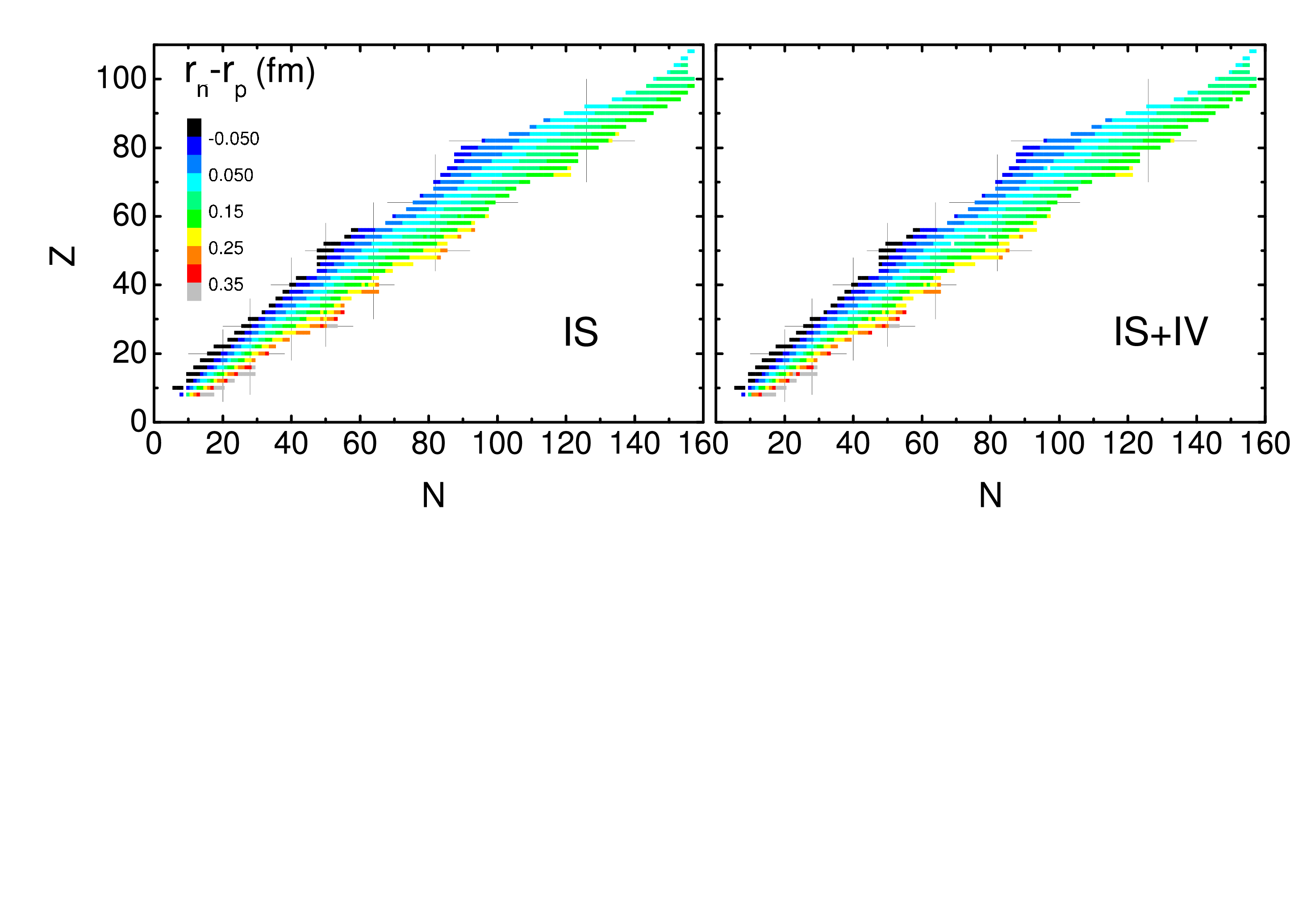}
\caption{(Color online) Neutron skin neutron  for isotopes and 
isotones  calculated
by means of the HF+BCS model with IS and IS+IV interactions. The thin  lines indicate 
the closed shells with N (or Z)=20, 28, 40, 50 ,64, 82 and 126. 
See the text for details. }
\label{n-skin}
\end{center}
\end{figure*}

 The average gaps $\Delta ^{(3)}$ are tabulated  for high and low isospins in 
Table \ref{table2}.  Each   isotope (isotone) in Fig. \ref{fig-Delta-np} is 
  divided into two subsets of almost equal numbers
of nuclei by a cut at some value of $I=(N-Z)/A$.  Both the average proton and neutron 
$\Delta ^{(3)}$ show smaller values for higher isospin so  that the pairing 
 interaction is weaker for neutron-rich nuclei.  The IS+IV interaction reproduces 
properly the difference of  the neutron $\Delta ^{(3)}$ between high and low isospin 
 nuclei.  For proton $\Delta ^{(3)}$ also, the IS+IV pairing gives a good account of the 
isospin effect than the IS pairing.  

\subsection{Nuclear radii}
Nuclear radii provide basic and important information for various aspect of nuclear 
structure problems.  The proton radii, or equivalently the charge radii with the correction of 
finite proton size, can be determined accurately by electron scattering and muon scattering 
experiments.  However it is difficult to determine the neutron radii of finite nuclei with the 
same accuracy level as that of the proton radii while there were several experimental 
attempts to determine the difference  of the neutron to proton 
 radius ~\cite{Sn-rnrp-exp,Trz01,rnrp-exp}.
It should be noticed that the difference 
of the neutron and proton radii,  $\delta r_{np}=r_n-r_p$, is called the neutron skin. It is thought that $\delta r_{np}$ can give 
important constrains on the effective interactions used in nuclear structure study \cite{Horwitz2001}.

The  neutron and proton radii of various isotopes and isotones are calculated by using the HF+BCS 
model with the two pairing interactions, IS and IS+IV.
The results of neutron radii are shown in the left panel of Fig. \ref{RnRp}.  
Since we do not find any appreciable 
 differences between the two pairing interactions in the results,
 the results of IS+IV interactions will be  mainly discussed  hereafter. 
   The results obtained by a
 simple empirical formula $r_n=r_0  N^{1/3}$ with $r_0=1.139$ fm  \cite{Meng06} 
  are also plotted in the figure. 
  In general,  the simple formula for $r_n$ agrees well with the HF+BCS results.  It is noticed that  
 the HF+BCS model gives larger neutron radii for  nuclei with $N<40$ than the simple formula but
 smaller for nuclei with $N>120$.    
The proton radii for $N=20$, 28, 40, 50, 82 and 126 isotones are shown as a function of 
proton number Z in the right panel.  The simple Z$^{1/3}$ dependence is also plotted to follow the formula 
 $r_p=1.263$/Z$^{1/3}$.  The simple formula in general gives a good account of the HF+BCS data 
 and could be 
a good starting point for describing the isospin dependence of nuclear charge radii.
 However we can see some deviation between the HF+BCS and the simple formula 
 especially heavy $N=50$ and $N=82$ isotones.

\begin{figure*}[htb]
\begin{center}
\includegraphics[clip,scale=0.5]{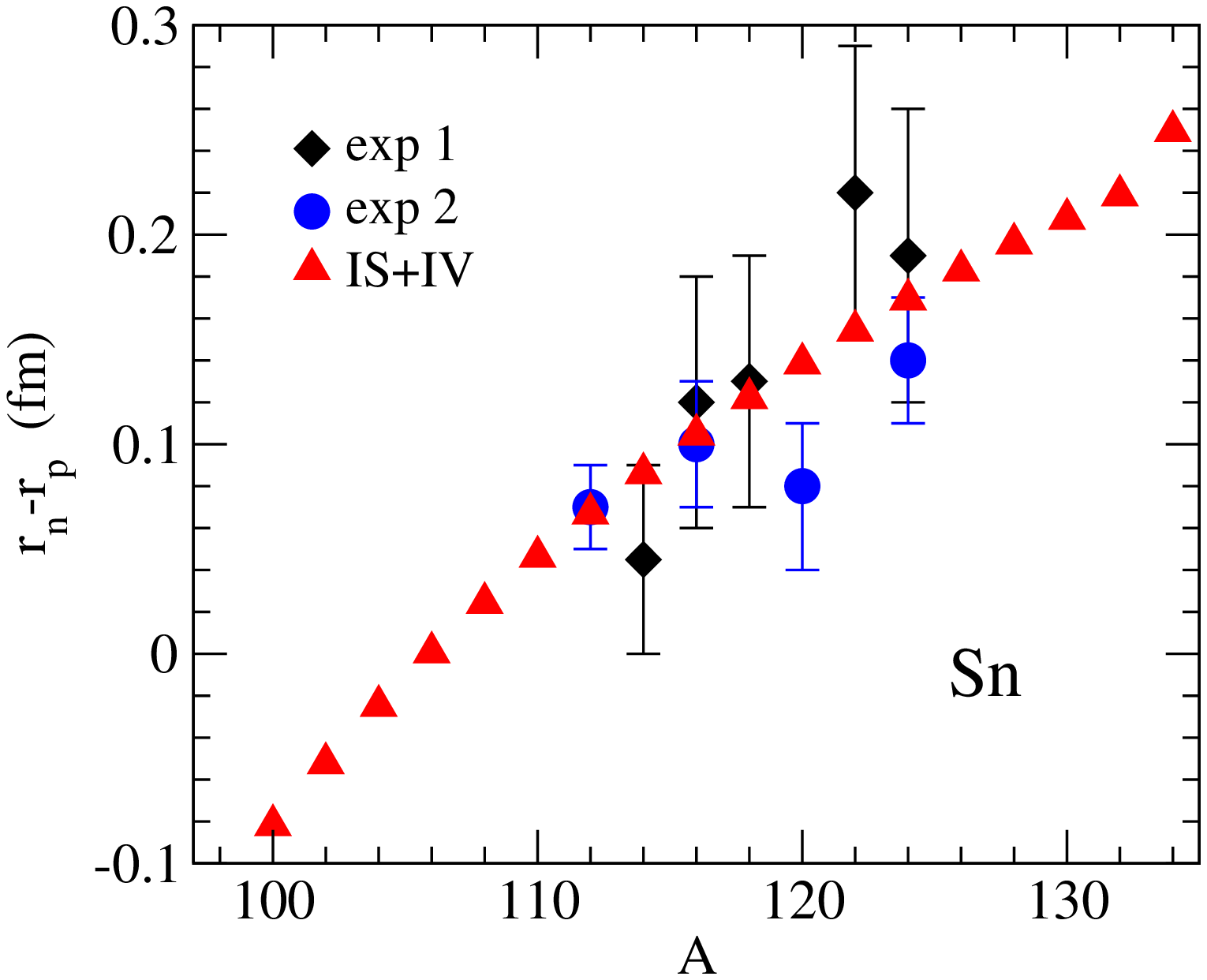}
\includegraphics[clip,scale=0.5]{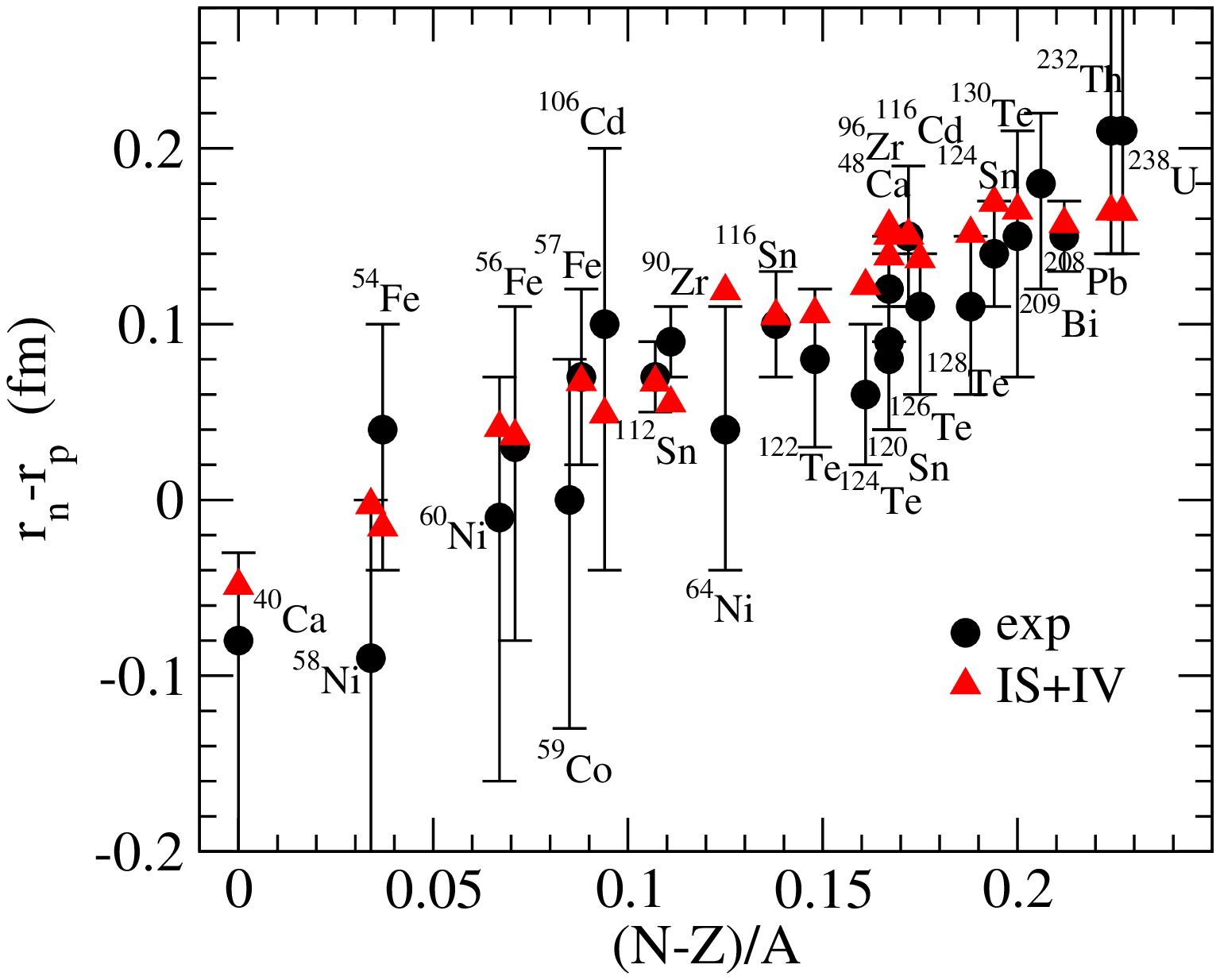}
\caption{(Color online) {\it Left} - Neutron skin  for Sn isotopes 
obtained with the HF+BCS model with IS+IV interaction.  
 Experimental data are taken from Ref. \cite{Sn-rnrp-exp,Trz01}.  
{\it Right} - Neutron skin as a function of isospin
parameter $I=(N-Z)/A$   calculated
by means of the  HF+BCS model with IS+IV interaction.  
 Experimental data are taken from Ref. \cite{rnrp-exp}.  
See the text for details.  }
\label{Sn-rnrp}
\end{center}
\end{figure*}

  The  neutron skin  $r_n-r_p$ calculated by HF+BCS model with the two pairing interactions
   are shown in Fig.  \ref{n-skin}.  The neutron skin becomes as large as 0.4 fm
 near the neutron drip line with $Z<28$. On the other hand, the neutron skin is at most 0.25 fm in
 neutron-rich nuclei with $Z>50$.  For proton-rich nuclei, the proton skin becomes 0.1 fm with $Z<56$
 and smaller than  0.05 fm  in heavier isotopes, larger than  $Z=56$.  The results of IS and IS+IV 
pairings are shown in the left panel and right panel, respectively. In general, the two 
pairing interactions give almost the same results as shown in Fig.  \ref{n-skin}.  However, it is noticed
 that the IS+IV pairing gives somewhat smaller neutron skins than 
the IS pairing  in very neutron-rich nuclei such as $^{136}$Sn, $^{150}$Ba and  $^{218}$Po.
 The calculated values are 
compared with empirical data of Sn isotopes obtained from studies of  spin-dipole resonances ~\cite{Sn-rnrp-exp} and
  antiprotonic atoms ~\cite{Trz01} in the left panel Fig. \ref{Sn-rnrp}.  
  The calculated values show reasonable agreement with the 
empirical data within the experimental error.  The isospin dependence of neutron skin is shown in 
 the right panel of  Fig. \ref{Sn-rnrp} 
 together with empirical values obtained by antiprotonic atom experiments in a wide range of  nuclei 
from $^{40}$Ca to $^{238}$U.   The slope of experimental data as a function of the isospin parameter 
 $I=(N-Z)/A$ is  reproduced well by our calculations.  
 
 The neutron skin of $^{208}$Pb has been 
 discussed intensively in relation with neutron matter properties.  The systematic studies of 
scattering data yield the empirical value  $r_n-r_p=0.17\pm 0.02$ fm which is close to another 
empirical value $r_n-r_p=0.15\pm0.02 $ fm from  the study of antiprotonic-atom systems.  
The model independent determination of parity violation experiment at Jefferson Laboratory ~\cite{Horwitz2001} 
has been proposed and performed recently to obtain the neutron skin of $^{208}$Pb. However 
 the statistics was poor and needs improvement by more data accumulation. 
Our calculated value  $r_n-r_p=0.157$ fm is close to the experimental values by the two systematic 
studies.

\section{Summary and conclusions}
In summary, we studied the binding energies, separation energies 
and OES  by using HF+BCS model with SLy4 
interactions together with the isospin dependence pairing (IS+IV
pairing)
 and isoscalar (IS pairing) interactions.  The calculations are performed with the EV8odd code for
even-even nuclei and also even-odd nuclei using the filling approximation.
  For the
 neutron pairing gaps,  the IS+IV  pairing strength
 decreases gradually as a function
 of the asymmetry parameter $(\rho_n(r)-\rho_p(r))/\rho(r)$.  On the other hand,
 the pairing strength for protons
increases for larger values of the asymmetry parameter
 because of the isospin factor in Eq. \eqref{eq:g1t}.
  The empirical isotope dependence of the neutron OES, $\Delta^{(3)}_n$, is well
  reproduced by the present calculations with the isospin dependent pairing
 compared with the IS pairing.   We can also obtain a good agreement between
 the experimental proton OES and the calculations with the
 isospin dependent pairing for $N=50$ and $N=82$ isotones.
 
The neutron and proton radii were also studied by using the same HF+BCS model with the two 
pairing interactions.  The two pairing interactions give essentially the same results for the radii 
 except for a few  very neutron-rich nuclei.  We found systematically large neutron skins in very 
neutron-rich nuclei with $|r_n-r_p|\sim0.4$ fm, while the proton skin is rather small even in nuclei 
close to the proton drip line because of the Coulomb interaction.  The calculated results of 
neutron skin show reasonable agreement with the empirical data including the ($N-Z$) dependence of
the data.

  We tested the IS+IV pairing for the Skyrme interaction SLy4  
  and found the results reproduce  well
 the systematical experimental data.
Thus, we confirm a  clear manifestation  of the isospin
 dependence of the   pairing interaction in the OES 
in comparison with the experimental data 
 both for protons and neutrons.
 
\begin{acknowledgments}
This work was partially supported  by the U.S. DOE
grants DE-FG02-08ER41533 and DE-FC02-07ER41457 (UNEDF, SciDAC-2),
the Research Corporation, and the JUSTIPEN/DOE grant DEFG02- 06ER41407, and
 by the Japanese Ministry of Education, Culture, Sports,
Science and Technology by Grant-in-Aid for Scientific Research under
the Program number C(2) 20540277. Computations were carried out on
the XT4 Jaguar supercomputer at the Oak Ridge National Laboratory.
\end{acknowledgments}

\end{document}